\documentclass{elsarticle}

\usepackage[hyphens]{url}
\usepackage{hyperref}

\usepackage{subfiles}
\usepackage{xcolor}
\usepackage{arydshln}
\usepackage{fancyvrb}
\usepackage{algorithm}
\usepackage[noend]{algpseudocode}

\usepackage{rotating}

\usepackage{booktabs}

\newcommand{\sslash}{\mathbin{/\mkern-5mu/}}

\algrenewcommand{\algorithmiccomment}[1]{\hfill{\tiny $\sslash$#1}}

\algnewcommand\algorithmicforeach{\textbf{for each}}
\algdef{S}[FOR]{ForEach}[1]{\algorithmicforeach\ #1\ \algorithmicdo}

\makeatletter
\def\algbackskip{\hskip-\ALG@thistlm}
\makeatother

\journal{Computers in Industry}

\begin{document}

\begin{frontmatter}



\title{Towards the implementation of Industry 4.0: a methodology-based approach oriented to the customer life cycle}

 \author{Víctor Julio Ramírez-Durán\corref{mycorrespondingauthor}}
 \cortext[mycorrespondingauthor]{Corresponding author}
 \ead{victorjulio.ramirez@ehu.eus}

 \author{Idoia Berges}
 \ead{idoia.berges@ehu.eus}

 \author{Arantza Illarramendi}
 \ead{a.illarramendi@ehu.eus}

 \address{Department of Languages and Information Systems, University of the Basque Country UPV/EHU, Donostia - San Sebastián, Spain, 20018}

\begin{abstract}
Many different worldwide initiatives are promoting the transformation from machine dominant manufacturing to digital manufacturing. Thus, to achieve a successful transformation to Industry 4.0 standard, manufacturing enterprises are required to implement a clear roadmap. However, Small and Medium Manufacturing Enterprises (SMEs) encounter many barriers and difficulties (economical, technical, cultural, etc.) in the implementation of Industry 4.0. Although several works deal with the incorporation of Industry 4.0 technologies in the area of the product and supply chain life cycles, which SMEs could use as reference, this is not the case for the customer life cycle. Thus, we present two contributions that can help the software engineers of those SMEs to incorporate Industry 4.0 technologies in the context of the customer life cycle. The first contribution is a methodology that can help those software engineers in the task of creating new software services, aligned with Industry 4.0, that allow to change how customers interact with enterprises and the experiences they have while interacting with them. The methodology details a set of stages that are divided into phases which in turn are made up of activities. It places special emphasis on the incorporation of semantics descriptions and 3D visualization in the implementation of those new services. The second contribution is a system developed for a real manufacturing scenario, using the proposed methodology, which allows to observe the possibilities that this kind of systems can offer to SMEs in two phases of the customer life cycle: Discover \& Shop, and Use \& Service. 

\end{abstract}


\end{frontmatter}



\section{Introduction}
\label{introduction}
The instantiation of the data-driven economy in the manufacturing industry has led to the development of 
different initiatives
and strategies addressing the use of data exploitation to optimize and transform the manufacturing business. Among those initiatives we can  find  ``Smart  Manufacturing" in USA, ``Made in CHINA 2025", ``Future Manufacturing" in UK and ``Industry 4.0" in Europe \cite{K2018}, which enable important business opportunities for the manufacturers.

Considering the Industry 4.0 in particular \cite{OG20}, this initiative  can have the greatest impact in three key areas regardless of the sector \cite{cotteleer_2017}: Products, supply chain, and customers. Regarding products, technologies such as sensors, machine learning or robotics can transform the way products are designed and developed, and the exploitation of data about the products can allow manufacturers to predict, plan and manage specific circumstances in order to optimize their production. Furthermore, new business models can be contemplated by selling data and services in addition to physical products. In the case of the supply chain, advanced forecasting techniques relying on internal (e.g. demand) and external (e.g. market trends) data can allow for a faster delivery time \cite{alicke_2016}. Moreover, real-time information about the supply network and the logistics capabilities can allow for a more flexible planning and inventory processes, which can react to changing demand or supply situations. Finally, regarding customers, Industry 4.0 technologies can help to gain a better understanding of the customers, enhance their experience when interacting with the products and enable better post-sale support. The proposal presented in this paper considers this last area.

However, what is observed is that most of research addressing implementation techniques in Industry 4.0 scenarios is created for large organizations and the majority of Small and Medium Enterprises (SMEs) are overwhelmed by the large amount of existing 4.0 technologies, the time and effort needed to learn them, and the costs of their implementation \cite{MS2020}. Moreover, while several works deal with the incorporation of Industry 4.0 in the area of the products (e.g. \cite{Tomic2020, MOURTZIS2018}) and the area of the supply chain (e.g. the survey in \cite{winkelhaus_2020}), literature on how to introduce Industry 4.0 technologies in the area of the customers is rather scarce. The proposal shown in this paper considers those two drawbacks: the difficulties of SMEs to achieve a successful transformation towards Industry 4.0 standard and the scare literature in the area of customer life cycle, and thus it  presents a new methodology which can help software engineers of SME manufacturing scenarios to incorporate Industry 4.0 technologies in the context of the customer life cycle. The goal pursued with this incorporation is to  improve efficiency and enhance customer experiences, thus helping manufactures to attract and retain customers.

The methodology consists of stages that are divided into phases which in turn are made up of activities. It  uses similar phases of those considered by a typical life cycle for the development of an information system \cite{Elmasri2010}, which are also used in well-known development models such as Waterfall, Iterative or Agile. But what makes it different and can be considered the main of our contributions are the detailed descriptions of the activities that must be carried out in each phase. 
Furthermore, one technical novelty  of the proposed  methodology  is the  commitment made in it to connect  semantic based technology, used to  represent knowledge related to manufactured products, with  3D digital technology, used to visualize  those products. Within semantic technologies, ontologies allow the representation of knowledge of a particular domain through the definition of categories, properties and relationships between concepts and entities in a way that is understandable by both machines and humans. The use of ontologies has been increasing over the years, making the knowledge represented through them much broader, facilitating the definition of new concepts through reuse and improving  interoperability between heterogeneous information
systems.  Furthermore, 3D rendering technologies allow detailed visualization of objects by adding navigation components that enhance the user experience. These technologies have evolved in such a way that those processes that some years ago took several minutes and needed a powerful system to be executed, nowadays can be carried out in milliseconds from the web browser of a conventional computer or from a mobile device (e.g. smartphones, tablets, laptops, etc.).
We believe that an accurate and detailed description of the products or services offered by the manufacturing industry together with an improved visual presentation, both enhanced by the use of these technologies, will enrich the experience offered to customers, positively affecting decision-making.

Although there exist other proposals in the smart manufacturing scenarios that advocate for the use of those technologies separately (e.g in \cite{EGOGEM19} authors propose a semantic rule language for industrial internet of things and in \cite{VANLOPIK2020} augmented reality technology  for industry 4.0), we have not found any other one that proposes the combined use of them. 


In order to show the feasibility of the methodology, its application has been deployed in a system for a real  manufacturing scenario in Urola Solutions\footnote{https://www.urolasolutions.com/en/}, a medium-sized enterprise located in the Basque Country and which belongs to the MONDRAGON Corporation\footnote{https://www.mondragon-corporation.com/en/}. Thus, the contribution of the paper is twofold: On the one hand, the methodology that guides software engineers of SMEs in the task of creating software services that allow to improve the relationship between customers and a company; and on the other hand, a system as a proof of concept, which allows to observe  the possibilities of that type of solutions. 


In the rest of the paper, a brief overview of the customer life cycle phases is presented first. Then, some related works to the use of semantic descriptions and 3D technologies in the manufacturing scenario are shown. Next, the proposed methodology with its stages, phases and  associated activities is described. Later,  an implementation of a case study for a real manufacturing scenario that considers two phases of the customer life cycle and which has been developed using the proposed methodology is introduced. Finally, some conclusions and future work are discussed.

\section{Brief Overview of the Customer Life Cycle}
The traditional customer life cycle consists of a series of five phases that the customer goes through on his way to acquire a good or service, with the objective of turning people into paying customers and achieving a loyalty relationship between the customer and the brand. These phases, which range from capturing the attention of a potential customer to achieving the aforementioned loyalty, are: Reach, Acquisition, Conversion, Retention and Loyalty. In the specific case of manufacturing, a framework for the customer life cycle oriented to Industry 4.0 is presented in \cite{hood2016industry}, which condenses the five general phases into three:
\begin{itemize}
    \item Discover \& shop: This phase has been greatly explored by e-commerce platforms such as Amazon\footnote{\url{https://www.amazon.com}},  AliExpress\footnote{\url{https://www.aliexpress.com}} or eBay\footnote{\url{https://www.ebay.com}}, mainly supported by the Business-to-Costumer (B2C) model, whose intention is to improve the end customer shopping experience by offering an intuitive experience and a recommendation system based on previous search and purchase preferences. However, in the manufacturing environment, and because of the complexity of their products, companies using the B2C model usually provide their customers with a generic physical or digital brochure with a lack of depth description of the offered goods or services.  This approach may end up generating more doubts in the customer than it intends to solve, thus requiring the intervention of a representative of the sales department in an additional process different from that which constitutes their task, which is to close the sale. Furthermore, these companies not only provide goods and services to end customers, but also other businesses, i.e. Business-to-Business (B2B) model, which means that, if the offer is not well detailed, the purchasing processes are carried out in extensive meetings between the purchasing and sales departments of each company, taking more time than actually necessary.
    \item Buy \& install: This phase covers the actual buying of the product by the end customer and its installation. Many manufacturers rely on channel partners, such as dealers or distributors, for managing these tasks, so it is important to guarantee a seamless and timely information sharing between the manufacturers and their channel partners. However, this is usually done through transactional and milestone-based systems intended to synchronize activities that often rely on manual inputs and updates or outdated mechanisms that run once a day \cite{hood2016industry}.
    \item Use \& service: This phase is oriented to customer retention and loyalty generation, and it is mainly applied in the B2B model, since the customer is tied to the use of products or services to manage their business. The most common way of providing a post-purchase service that includes repair, maintenance and support is through the customer service lines. However, this solution presents different problems such as fixed service hours, congestion on the lines, relatively long waiting times and the high possibility of not finding an immediate solution. These problems generate high customer discomfort creating an effect contrary to the desired loyalty. The effect is much more damaging in the manufacturing industry, where enormous economic loss can be generated by not having the requested spare part or not having carried out preventive and corrective maintenance on time.
\end{itemize}

In the case study presented in this paper two of the three phases presented above are considered: Discover \& Shop, and  Use \& Service. Regarding Discover \& Shop two services have been implemented. The first one, allows customers to approach to the products in which they are interested by  filling out a simple questionnaire and the second one allows customers to navigate through those products. The service implemented in relation  to Use \& Service is a kind of virtual technician that tries to solve the needs in terms of requesting spare parts.

\section{Related Work}
\label{relatedwork}

There exist several technologies that enable the implementation of Industry 4.0, such as Internet of Things (IoT), Cloud Computing, Cybersecurity, Big Data and Analytics, Augmented/Virtual Reality, Additive Manufacturing, Simulation, and Robotics. In the specialized literature, several projects and case studies using these technologies in the context of manufacturing SMEs can be found. For example, project ESMERA (European SMEs Robotic Applications \cite{Icer2018}) of the European Commission’s Horizon 2020 Research and Innovation Programme aims to boost robotics innovation for European SMEs by funding projects such as REFLECT \cite{reflect_2020}, which tackles the assembly of deformable parts in dishwashers by using a robotic system. Also under the Horizon 2020 Programme, project CloudiFacturing \cite{cloudifacturing_2020} aims to optimize production processes and producibility in SMEs using Cloud/HPC-based modeling and simulation. More precisely, it supports projects such as 3D-CPAM (3D Clothing Production by Additive Manufacturing \cite{Tomic2020}), which uses advanced HPC/Cloud services and modern 3D printing technologies to optimize the 3D fashion design manufacturing process, or D2Twin \cite{d2twin_2020}, which uses big data analytics to improve quality control and maintenance.

As mentioned in the previous section, our proposal is supported by two pillars: semantic-based technology to represent knowledge related to manufactured products, and 3D digital technology to visualize those products.
These pillars are highly related to two of the aforementioned Industry 4.0 enablers: IoT and Augmented/Virtual reality.

From the IoT perspective where equipment and products communicate and are connected to each other, interoperating and integrating data 
and information is a demanding task that can be facilitated by semantic technologies such as ontologies. Ontologies allow to represent the semantics of knowledge  and data in a formal, comprehensive and reusable way. Thus, several ontologies with different purposes have been defined in manufacturing scenarios, such as:
the PSL ontology \cite{Gruninger09}, which includes fundamental concepts for representing manufacturing processes; the MASON ontology \cite{Lemaignan06}, an upper ontology that represents what authors consider the core concepts of the manufacturing domain: products, processes and resources; the SIMPM  ontology \cite{Sormaz2019}, an upper ontology that models the fundamental constraints of manufacturing process planning: manufacturing activities and resources, time and aggregation; the MaRCO  ontology \cite{Jarvenpaa2019}, which defines capabilities of manufacturing resources; the MSDL ontology \cite{ameri2006upper}, which allows to describe manufacturing services; the P-PSO ontology \cite{Garetti12}, which considers three aspects in the manufacturing domain: the physical aspect (the material definition of the system), the technological aspect (the operational view of the system) and the control aspect (the management activities), for information exchange, design, control, simulation and other applications;  OntoSTEP  \cite{OntoStep12}, which allows the description of product information mainly related to geometry; MCCO  \cite{MCCO}, which  focuses on interoperability across the production and design domains of product lifecycle; ExtruOnt\cite{ramirez2020}, which describes different aspects of extrusion machines such as their components and the 3D position, the spatial connections and the features of those components; CMO \cite{Talhi2019}, which represents the cloud manufacturing domain to support information exchange between cloud manufacturing resources; and SAREF4INMA \cite{deRoode2020}, which pursues favouring interoperability with industry standards. Finally, a literature review of papers related to ontologies in the area of product lifecycle management is presented in \cite{Fortineau2013}.

Regarding 3D digital technology, Virtual Reality (VR), Augmented Reality (AR) and Mixed Reality (MR) are considered relevant technologies for the new generation of intelligent manufacturing \cite{ZZRL19}. Virtual reality is a high-end human-computer interface that allows interaction with simulated environments in real time and through multiple sensorial channels \cite{LIAGKOU2019}. The users believe to be inside a reality that does not exist in truth, but they act like in the real world \cite{S09}. Augmented Reality  has been defined as a system which supplements the real world with virtual objects (computer generated)  that appear to coexist in the same space as the real world \cite {BV19, MASOOD2019}. It provides benefits especially in designing products and production systems. While VR requires inhabiting an entirely virtual environment, AR uses existing natural environment and overlays virtual information on top of it.  Finally, Mixed Reality like augmented reality, places digital or virtual objects in the real world. However, with mixed reality, users can quickly and easily interact with those digital objects to enhance their experience of reality or improve efficiency with certain tasks \cite{GPCL17}. 
These technologies can been used for several purposes in the industrial and manufacturing environment, for example in the process of product design \cite{MOURTZIS2018, Guo2018}, for assembly simulations \cite{Tao2021, Al-Ahmari2016}, for training purposes \cite{Ordaz2015, Tao2019}, for factory layout planning \cite{Gong2019, Herr2018} or for improving maintenance services \cite{Riboldi2021, Ababsa2020}. Moreover, since 3D modeling has shown a realistic description  of manufactured products by generating high-quality textures and proper lighting, it can be used for showcasing purposes. In this sense, it can be seen how some companies such as Schneider Electric\footnote{\url{https://sketchfab.com/blogs/enterprise/customer-stories/electronics/3d-viewer-schneider-electric}} or Reid Supply\footnote{\url{https://www.reidsupply.com/}} show their technical products in a interactive 3D product catalogue.


Nevertheless, to the best of our knowledge, no proposal has been made which combines the use of 3D and semantic technologies for a customer life cycle framework in a smart manufacturing scenario. Thus, in this paper, an initiative of this type is proposed.


\section{Methodology}
\begin{sidewaysfigure} 
\centering
\includegraphics[width=\columnwidth]{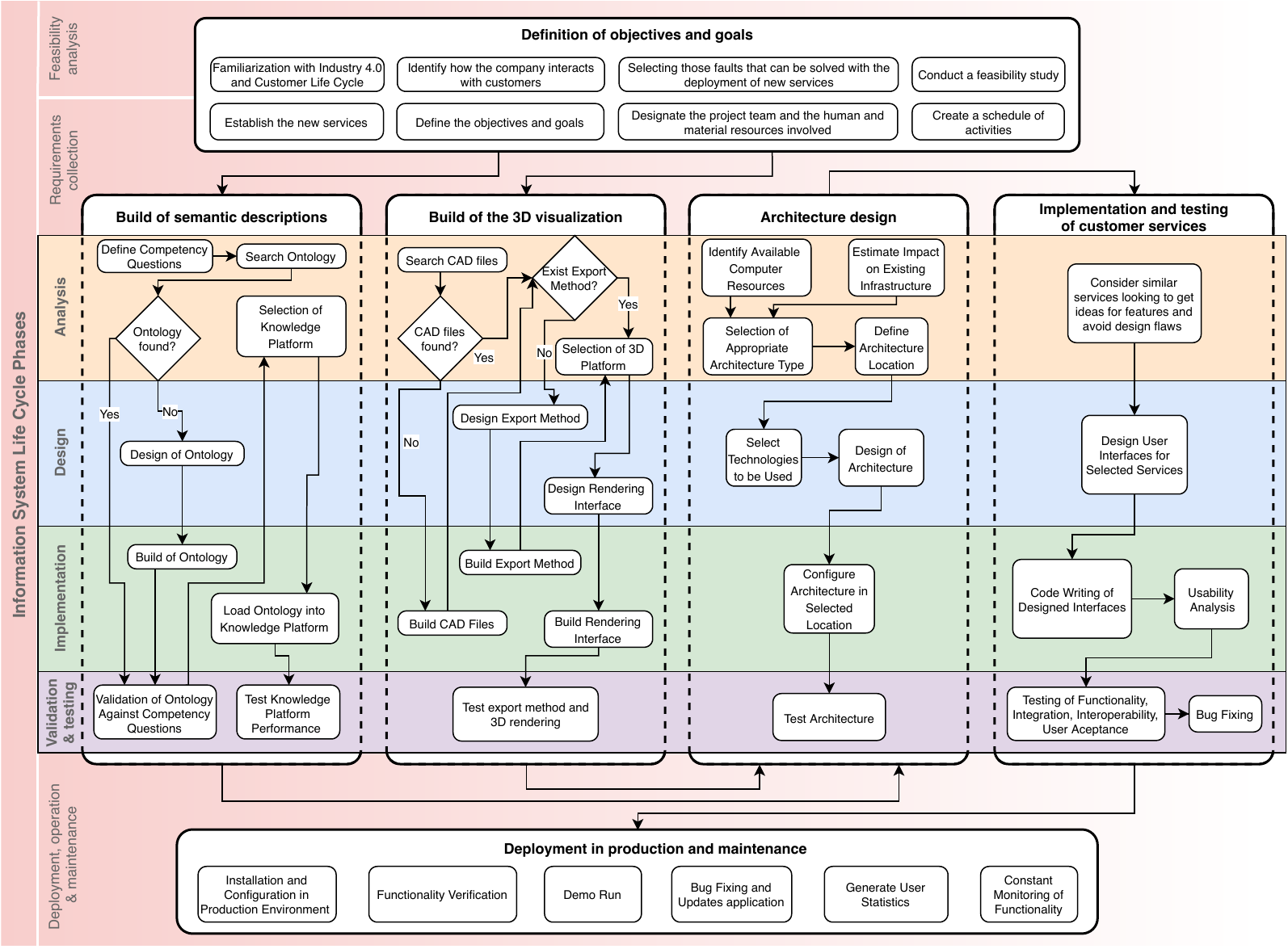}
  \caption{\label{fig:methodology}
     Methodology diagram.}
      \end{sidewaysfigure}

The approach  introduced in this paper is intended to illustrate how SMEs could  achieve closer and long-lasting  relationships with customers through the implementation of new services aligned with the four industrial revolution (Industry 4.0). In this line, we present a step-by-step methodology that can help technicians of those SMEs in the task of deploying those services.


The proposed methodology considers the phases of a typical life cycle of an information system \cite{Elmasri2010} but  what makes it different are the detailed guides that it provides for the incorporation of semantics descriptions and 3D visualization in the implementation of services associated to the customer life cycle. The methodology consists of stages that are divided into phases which in turn are made up of activities  (see figure \ref{fig:methodology}).
In table \ref{table:inputsOutputs}, a summary of the inputs and outputs of the methodology stages is presented. It represents the necessary income to obtain the desired result through the execution of the activities described in each stage of the proposed methodology. The different stages and activities that make up the proposed methodology are presented below.


\begin{table}[]
    \centering
    \resizebox{\linewidth}{!}{%
    \begin{tabular}{lll}
    \toprule
        \textbf{Stage} & \textbf{Input} & \textbf{Output} \\
    \midrule
         Definition of objectives and goals & Interest of adopting Industry 4.0  & List of new services, objectives, goals\\
          & for improving relationships with &  and roadmap for the deployment \\
          & customers & of those services \\
          \hdashline
         Build of semantic descriptions & Specifications of the manufactured & Selected ontology and \\
         & products & knowledge platform \\
          \hdashline
         Build of the 3D visualization & List of manufactured products &  Export/synchronization  \\
          &  & method and rendering interface \\
          \hdashline
         Architecture design & Quality attribute requirements, & An architecture\\
          & knowledge platform, export method  & \\
          & and rendering interface & \\
          \hdashline
         Implementation and testing  & Definition of customer services & Implemented customer services \\
          of customer services & & \\
          \hdashline
         Deployment in production and & Implemented customer services & Customer services deployed in \\
          maintenance & & production and maintenance \\
          & & period started\\
    \bottomrule
    \end{tabular}}
    \caption{Inputs and outputs of methodology stages}
    \label{table:inputsOutputs}
\end{table}

\subsection{Definition of objectives and goals}

Based on the interest of the company in adopting Industry 4.0 for improving relationships with customers, the aim of this stage is to define a roadmap for the deployment of new services that can overcome some of identified weaknesses. Thus, the first activity is to carry out a \textit{familiarization with Industry 4.0 and Customer Life Cycle}, where Industry 4.0 benefits and customer life cycle phases are presented in such a way that the concepts are clear and familiar to stakeholders.


The next activity is to \textit{identify how the company interacts with customers} with the purpose of detecting  problems and limitations.  It is important to analyze the strengths and disadvantages of the business model in relation to competitors, and also  the statistics of the customer support department, for example, to see which are the most frequent customer complaints.

Once the problems and limitations have been identified, it is necessary to narrow the scope by  \textit{selecting those faults that can be solved with the deployment of services aligned with the  Industry 4.0}. Those services could include product exploration, training, maintenance, etc. in such a way that they allow to improve the relationships between customers and the company. This selection will be much more accurate thanks to the familiarization with Industry 4.0 and customer life cycle made previously.

The next activity is to \textit{establish those new  services} to be implemented throughout the customer life cycle. The \textit{objectives and goals} (along with their priorities), that have to be achieved with the implementation of those services, must also be defined followed by a \textit{feasibility study} that demonstrates that the project is profitable and that the company has the technical and organizational capacities to carry it out.
In order to complete the roadmap, it is also important to \textit{designate the project team and the human and material resources involved} as well as to \textit{create a schedule of the activities} necessary to implement the project, including those responsible of each of them.

\subsection{Build of semantic descriptions}

One of the two pillars on which the proposed methodology is based on is the use of semantic descriptions for representing knowledge related to manufactured products. 
The detailed semantic description of the products and services will guarantee the creation of, for example, a robust catalogue, an accurate search engine and an improved technical support. 
To accomplish this, the first activity is to \textit{define the ontology competency questions}. Competency Questions (CQs) are natural language questions outlining and constraining the scope of knowledge represented in an ontology \cite{WISNIEWSKI2019100534}. Within the framework of this methodology, the competency questions should be aimed at describing the company's products and services, in addition to answering those questions that the customer may ask about their properties or characteristics.

Next, it is required to \textit{search an ontology} that correctly answers those defined competency questions and, if necessary, make the pertinent modifications to adapt it correctly to what it has been looking for. There are different repositories where ontologies that span multiple domains can be found, such as LOV \cite{LOV2017}, Swoogle \cite{swoogle2004}, ODP \cite{gangemi2010ontology} and Ontohub \cite{codescu2017ontohub}, however, there are also ontologies created as result of scientific research and that are not found in those repositories, hence a search using the most popular search engines is also recommended.
In the event that it is not possible to find an ontology that meets all the requirements, it is necessary to \textit{design the ontology}. In the literature different methodologies such as On-To-Knowledge \cite{Sure2004}, Diligent \cite{Pinto2004} and NeOn \cite{Neon2012} can be found to adequately guide engineers through a step-by-step ontology development process. 
Once the ontology is designed, the \textit{ontology must be built}.
This activity can be done making use of free tools\footnote{A complete listing of these tools can be found at \url{https://www.w3.org/wiki/Ontology_editors}.} such as Protegé \cite{noy2003protege}, NeOn Toolkit \cite{Erdmann2012} and SWOOP \cite{KALYANPUR2006144}, which facilitate creation and editing, as well as verification of its structure and integrity.
Finally, an \textit{evaluation of the selected or built ontology against the competency questions} must be carried out in order to validate that they can all be answered. If this is not the case, it is necessary to go back to the search/design activity.

One might argue that an underlying ontology could be disregarded in favour of a database. However, the degree of flexibility that ontologies offer for representing the hierarchy and properties of individuals, as well as querying about them, cannot be achieved by databases. Moreover, ontologies are better suited for dealing with inheritance and are prepared for reasoning and inference purposes. In addition, there is usually a looser coupling between ontologies and applications that use them than between database schemas and applications. Finally, ontologies preserve semantics and provide a shared understanding of a domain, favouring interoperability \cite{uschold_15}.

With the ontology finally defined, it is time to \textit{select the knowledge platform} that will storage and sustain the ontology. Ontology storage models are classified into native stores and database stores. Native stores are directly built on the file system, whereas database-based repositories use relational or object relational databases as backend store. Storing ontologies using a native storage solution is straightforward compared to storing ontologies in databases, as relational databases do not support hierarchical relations directly \cite{Abburu2016}. 
Although the selection of the most suitable knowledge platform to store the ontology depends on several performance factors such as the response times of the queries and the speed of data loading, there are evaluations such as those presented in \cite{addlesee_2019,addlesee_2019b} which help facilitate the process of selection.
There are multiple ontology stores based on different technologies, some of which offer a free version with most of their functionalities. Examples of these are Virtuoso\footnote{\url{http://vos.openlinksw.com/owiki/wiki/VOS}}, Stardog\footnote{\url{https://www.stardog.com/}}, RDFox\footnote{\url{https://www.oxfordsemantic.tech/product}}, GraphDB\footnote{\url{http://graphdb.ontotext.com/}}, Blazegraph\footnote{\url{https://blazegraph.com/}} and AnzoGraph\footnote{\url{https://www.cambridgesemantics.com/anzograph/}}. 

The last two activities related to the build of semantic descriptions are the \textit{loading of the ontology in the selected knowledge platform} and the \textit{performance test of the knowledge platform}. In these activities, it is important to verify that there are no incompatibilities between the ontology and the selected knowledge platform, the data can be loaded and queried without problems, the necessary endpoints are provided and the performance is adequate. This can be done by uploading synthetic product and service data to the knowledge platform, converting the competency questions to the required query language, and executing those queries.

\subsection{Build of the 3D visualization}
\label{section:build3dvisualization}
The second pillar on which the proposed methodology is based on is the use of 3D rendering technologies for the visualization of products in order to provide an enhanced experience to customers.
For this, it is necessary at this stage to verify the existence of the 3D models of the products and define how those models will be visualized by the customers. Furthermore, it is also necessary to establish a synchronization method between the modifications made to the 3D models of products and their final presentation in such a way that the changes made in the design are reflected in the customer's visualization.

The first activity consists of \textit{verifying if there is a 3D representation (CAD files)} of the products offered by the company. Many manufacturing enterprises have those representations in the form of CAD files, which features can vary depending on the software used to create them. 
In the event that the company does not have 3D representations of its products, it is possible to search for them in
free repositories like 3D Warehouse\footnote{\url{https://3dwarehouse.sketchup.com/}}, GrabCAD\footnote{\url{https://grabcad.com/}} and traceparts\footnote{\url{https://www.traceparts.com/}}, and other paid ones such as Sketchfab\footnote{\url{https://sketchfab.com/}} where it can be found thousand of 3D models divided by categories.

If product CAD files are not available, \textit{they must be built} using a CAD software. For this purpose, there exist free alternatives for desktop (e.g. FreeCAD\footnote{\url{https://www.freecadweb.org/}}) or web environments (e.g. Onshape\footnote{\url{https://www.onshape.com/}}, Autodesk Fusion 360\footnote{\url{https://www.autodesk.es/products/fusion-360/subscribe}}, CMS IntelliCAD PE\footnote{\url{https://www.intellicadms.com/}}). Some of the latter are based on cloud processing, which introduces some key advantages such as that there is no need to invest in computer resources for processing, the latest version is always available, new features and bug fixes are added automatically, can be accessed from anywhere and can cost up to a third of the value of traditional software\footnote{\url{https://3dstartpoint.com/cloud-based-cad-software-101-how-it-works-and-top-5-picks/}}.

The next activity is to \textit{verify if an export/synchronization method exists}, in such a way that the changes made in the design are visualized by the customer. Most CAD software provide a manual option to export 3D models to local disk in different formats such as X3D, STL, GLTF or OBJ. However, it must be established if there is an automated option in which 3D models can be transferred to the server from where they will be rendered. If this option is not available, it is necessary to\textit{ design and build a synchronization method}. There are manual and automated methods to carry out the synchronization. The most basic one suggests that the designer export the modification and load it into the repository used by the web application. This method adds an extra burden to the designer's work and is prone to desynchronizations. Another approach suggests using CRON tasks that run the synchronization process on a set schedule or at times of low network traffic. It is also possible to synchronize using triggers at the time the designer makes a modification or the client visualizes a product, for which it is necessary to develop a query method to ask for recent modifications, either through an API in the client application or in the selected design software.

Once the export/synchronization method has been defined, \textit{the 3D platform must be selected}, that is, the technology in charge of rendering the CAD models in the web environment. There are several free libraries and frameworks based on WebGL (an OpenGL based javascript library) for 3D rendering in web browsers\footnote{A listing of the available frameworks (free and paid) for creating 3D content can be found at \url{https://en.wikipedia.org/wiki/List_of_WebGL_frameworks}} such as: Three.js\footnote{\url{https://threejs.org/}}, Babylon.js\footnote{\url{https://www.babylonjs.com/}} and X3DOM\footnote{\url{https://www.x3dom.org/}}. Selection of the most appropriate should be carried out taking into account previous experience, visualization needs and learning curve.

Next, the \textit{design and build of the rendering interface} must be carried out. This consists of not only being able to render 3D models on screen, but also to offer the necessary navigation and exploration tools to improve the customer experience. For this purpose, it is necessary that the designed rendering interface allows, in addition to rendering the 3D models optimally and appropriately, modifying the visualization to add relevant information as a result of the annotations created in the ontology (name, model and characteristics of parts, etc.).
The final activity consists of \textit{performing communication tests between the CAD software and the rendering interface} using the export/synchronization method in order to check the compatibility, synchronization and visualization of the exported CAD models.

\subsection{Architecture design}

At this stage, the architecture that should support the knowledge platform, the export/synchronization method and the rendering interface must be defined. Moreover, as it is a solution oriented to the customer life cycle, it is necessary to define an architecture capable of supporting all the necessary services (customer management, catalogue, search engine, technical support, etc.).

Before selecting the type of architecture to use, it is necessary to \textit{identify the available computing resources} as well as to \textit{estimate the impact that the implementation of the new architecture} will have on the existing infrastructure. That is, to verify that the requirements for the inclusion of new technologies do not conflict with the requirements for the correct execution of existing applications (i.e. operating system and libraries versions) and that the current computing resources are sufficient or it is necessary to incur in an economic investment.

The next activity is to \textit{select the appropriate architecture type}. Among the architectures for distributed systems \cite{puder2011distributed} the best known are client/server, where clients contact the server, which is responsible for handling requests; n-tier, where (Web) clients interact with front-end services that then delegate requests to their (database) back ends; peer-to-peer, where each and every node can do both request and respond for the services; and Service-Oriented (SOA), where services are provided to the other components by application components, through a communication protocol over a network. In the framework of the Industry 4.0, several service-oriented architectures has been proposed such as RAMI4.0 \cite{schweichhart2016reference}, ARUM \cite{Leito2013MultiagentSA}, SOCRADES \cite{KARNOUSKOS20092113}, PERFoRM, IMPROVE and BaSys4.0 \cite{trunzer2019system}. The selection of the architecture type should be based on the previous two activities and on the projections that the company has for its current infrastructure. The \textit{definition of the architecture location} also depends on the latter, that is, if the architecture will be implemented in a client's own infrastructure or in the cloud. Currently, there are different cloud service providers (e.g. Amazon AWS\footnote{\url{https://aws.amazon.com/}}, Google Cloud\footnote{\url{https://cloud.google.com/}}, Microsoft Azure\footnote{\url{https://azure.microsoft.com/}}) which allow the implementation of infrastructure in an agile, configurable and secure way for a monthly cost, avoiding the need to incur in a high immediate economic expense.

The next activity is to \textit{select the technologies that will be used in the design of the architecture}. Different technologies can be used for communication from the client application to the database server, such as web services or RESTful API's. The use of these technologies is recommended since it facilitates scalability and maintenance in the case of multiple client applications. Next, the \textit{architecture design} must be carried out, which is essential to develop and maintain large-scale, long-living software systems. As stated in \cite{hasselbring2018software}, the architecture defines the system in terms of components and connections among those components. Moreover, the architecture shows the correspondence between the requirements and the constructed system, thereby providing some rationale for the design decisions. The architecture must be designed keeping in mind all the quality attribute requirements specified for the project, such as performance and reliability.

The last two activities consist of \textit{configuring the architecture in the selected location} and \textit{testing the architecture}. This last activity can be performed by defining use cases that describe specific interactions between the user of the system and the system itself, and testing those use cases against the implemented architecture checking its stability, performance, security and reliability.


\subsection{Implementation and testing of customer services}
In this stage, the implementation and testing of the services defined in the \textit{Definition of objectives and goals} stage is carried out. These services must be aimed at guaranteeing an improvement in the customer life cycle through the incorporation of the technologies selected in the previous stages.

At first, it is important to \textit{consider similar services} to those to be implemented, so that ideas can be obtained on how to add features and avoid design problems. That is, if one of the services to be created is a product catalogue, it is advisable to review industrial catalogues, such as Schneider EZList\footnote{\url{https://ezlist.schneider-electric.com/}} and Siemens Industry Mall\footnote{\url{https://mall.industry.siemens.com/}}, and popular online stores such as Amazon, AliExpress and eBay, looking for insights about the distribution of the sections on screen or, if a search service is included, services like Google or Yahoo can be checked to identify important functionalities (e.g. autocomplete, suggestions).

Next, the \textit{design of user interfaces for the selected services} must be done. This process is not trivial as, in multiple times, users have to deal with frustration, fear and failure when faced with overly complex menus, incomprehensible terminology or chaotic navigation routes. To address this, interfaces should reduce anxiety and fear of use (embarrassing mistakes, privacy breaches, fear of scams), allow a smooth evolution (transition from novice to expert), allow compatibility with different input devices ( keyboard, mouse, multi-touch displays, gestural input, haptic devices, VR devices), provide online help (text, video tutorials), improve the exploration of information (filter, select, navigate with minimum effort and without fear of getting lost). Designers should start by: 1) determining user needs: a thoroughly documented set of user needs clarifies the design process; 2) generating multiple design alternatives: rethinking interface designs for different situations often results in a better product for all users; and 3) carrying out extensive evaluations: which can be done before fully programming the functionality of the interfaces using sketches. Low-fidelity paper sketches are helpful, but online high-fidelity prototypes create a more realistic environment for expert review \cite{Shneiderman2016}.

The next activity consists of \textit{code writing the services using the designed interfaces} followed by an \textit{usability analysis}. This activity must be carried out using the software tools and programming languages preferred by each company according to its infrastructure and experience. This task should be modest if the interface design is complete and accurate. The usability analysis must be designed to find flaws in the developed services taking into account various evaluation criteria such as: time to learn, speed of performance, rate of errors by users, retention over time and subjective satisfaction. The language and expressions used in the developed services must also be taken into account. If the company operates internationally, translation into multiple languages is necessary, otherwise it must be adapted to the local culture. In \cite{Bevan2016} several ISO standards related to usability are presented that serve as a guide to carry out this analysis: definition and concepts, evaluation reports, quality metrics, etc.

Finally, it must proceed with the \textit{functionality, integration, interoperability and acceptance tests}. Those tests certify that the developed services meets the goals of designers and customers, moreover, a carefully tested prototype generates little change during deployment, avoiding costly upgrades. In those tests, a set of use cases for the services must be specified, with defined requirements such as the minimum response time for the combination of software and hardware or the degree of user acceptance measured through satisfaction surveys. If the services fail to meet these criteria, the services must be reworked. This testing activity usually results in a large number of bugs to be fixed which demands the use of many human resources, making the \textit{bug fixing} a disorganized, chaotic and time-consuming process. The use of bug-tracking systems (e.g. Monday\footnote{\url{https://monday.com/s/bug-tracking-software/}}, Airbrake\footnote{\url{https://airbrake.io/}}, Backlog\footnote{\url{https://backlog.com/bug-tracking-software/}}, Bugzilla\footnote{\url{https://www.bugzilla.org/about/}}, Mantis\footnote{\url{https://www.mantisbt.org/}}) facilitates this activity, assigning a state to each detected bug, managing the available resources and maintaining a traceability over the process.

\subsection{Deployment in production and maintenance}

In this last stage, the \textit{developed services are installed and configured in the production environment}. 
It is recommended to use a version control in such a way that from now on all the changes made have a traceability. Version control helps teams keep track of all individual changes and prevent concurrent work from conflicting. In \cite{Rao2016}, a comparison of different version control systems can be found, which can help to select the one that best suits the needs of the company.

Once the services have been installed and configured, a \textit{verification of their functionality} must be carried out to ensure that there are no incompatibilities with the infrastructure available in production. Furthermore, a \textit{demo with real users must be carried out} in order to certify that the services behave the way they were designed. If there are errors detected in the production environment, \textit{they must be corrected and updates to the libraries and packages} that are out of date in this environment, \textit{must be performed}. Once the operation of the services is validated and approved, the maintenance period begins. During this period, the \textit{functionality of the services must be constantly monitored} and \textit{statistics on their use must be generated}, in order to assess whether the services positively affected customers throughout their life cycle.



\section{Case Study}
In this section we present a case study in which the methodology described in the previous section has been applied to develop three services that allow to improve the customer life cycle in a real manufacturing company: Urola Solutions. That company develops advanced solutions for the packaging manufacture using blow moulding technology. Among the products they offer, there are different types of extruders depending on their production capacity and the types of plastic they use for packaging.

As stated in the first stage of the proposed methodology, first of all an analysis of the type of interaction between the company and its customers must be carried out in order to detect those faults that can be solved with the deployment of new services.

\subsection{Definition of objectives and goals}

Urola Solutions has a fairly solid range of clients and the acquisition of new clients is mainly based on the reputation achieved over time and the good references obtained as a result of these consolidated relationships.
However, regarding potential new customers  who browse through its website, currently they only see a general description of the products and services offered, in addition to the form and the telephone numbers of the contact section. As a consequence, requests for information are very common, and thus a personalized attention at the first contact between the future customer and the sales representative is an excessive burden for the company. A  similar situation occurs with the customer service provided, where for a remote assistance an initial survey must be carried out before being redirected to the suitable specialist, increasing occupation times of customer service representatives and congestion on the lines.

Taking into account the aforementioned faults and a previous meeting where the phases of the customer life cycle and Industry 4.0 implementation advantages were exposed, the main goal of this case study was oriented to improve customers experience in two phases of the customer life cycle, Discover \& Shop  and Use \& Service, through the development of three services: catalogue, searching module and virtual technician.
For this purpose, the following general objectives were defined:

\begin{itemize}
    \item Discover \& Shop
    \begin{itemize}
        \item Creation of a searching module with questions in natural language that allows customers to find the right product according to their production needs in a simple and efficient way.
        \item Creation of an improved catalogue of the company's products, with 3D visualization and advanced navigation options that allows customers to see those products in detail.
    \end{itemize}
    \item Use \& Service
    \begin{itemize}
        \item Creation of a virtual technician, which will try to solve the needs in terms of requesting spare parts and will only redirect the most complex cases to the appropriate specialist.
    \end{itemize}
\end{itemize}

A feasibility study was conducted to determine if the company had the necessary staff, technical resources and capital to develop the project\footnote{The details of the feasibility study are not available due to the company's data privacy policies}, and as a result some guidelines were established: the project duration could not be longer that 8 months and no more than 3 company employees should be involved with a weekly dedication of 32 hours each as a maximum. Considering those guidelines, a team of 6 members was assembled as follow: two software engineers, one for the backend (API, Database) and another for the frontend (Application, Users management), with previous experience for the necessary programming and the R\&D manager (with direction role) regarding company employees; and the three authors of this paper with guidance and support roles. The design of service interfaces was supported by a company designer with no direct commitment to the project.

\begin{sidewaysfigure} 
\centering
\includegraphics[width=\columnwidth]{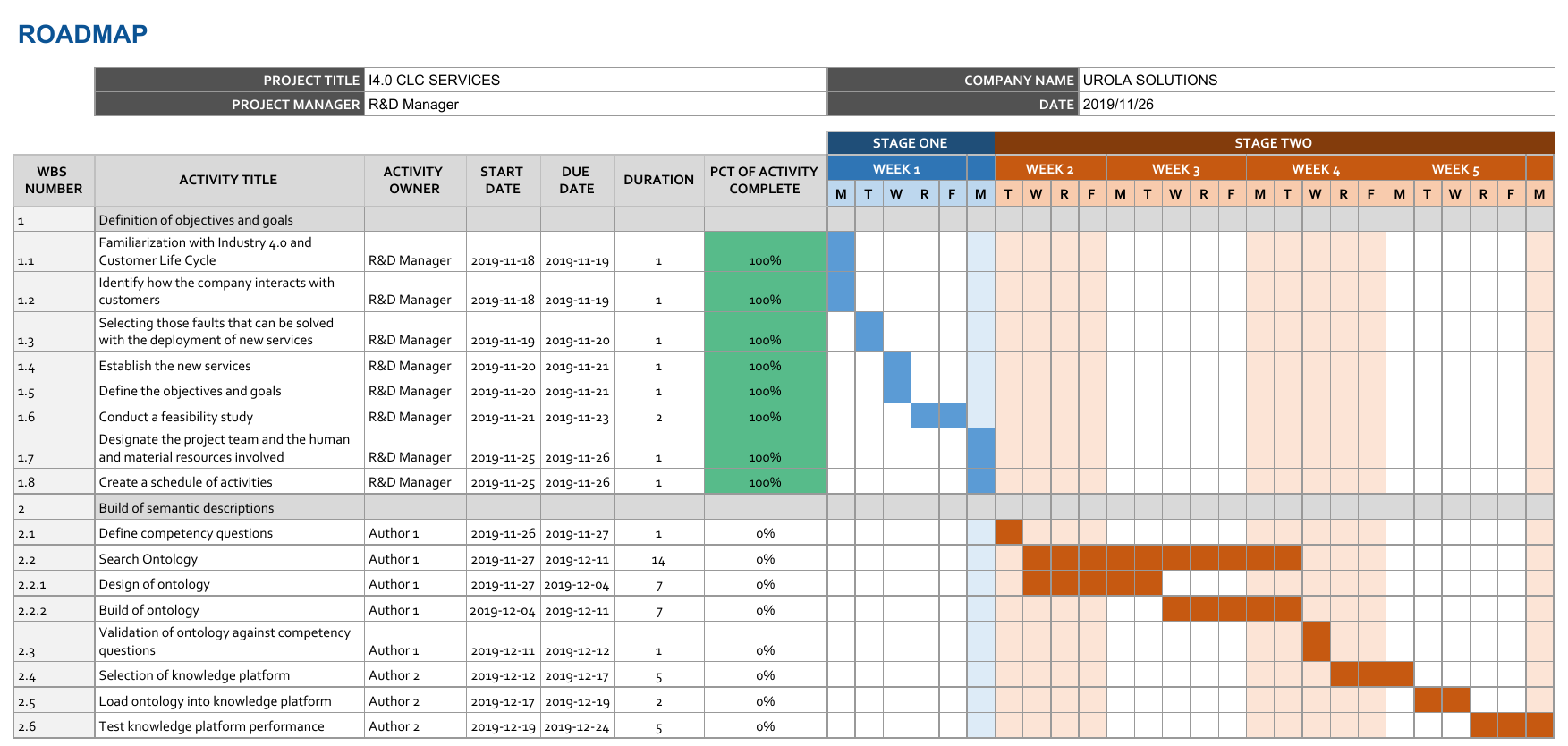}
  \caption{\label{fig:roadmap}
     Excerpt of the project roadmap.}
      \end{sidewaysfigure}

The duration of stages in Table \ref{table:schedule} was estimated for a period of 6 months taking into account the next stages of the proposed methodology. Moreover, in figure \ref{fig:roadmap} an excerpt of the project roadmap defined at that time using the activities indicated in the proposed methodology is shown.

\begin{table}[h!]
\centering
\begin{tabular}{l c} 
 \hline
 Stage & \# of Weeks \\ 
 \hline\hline
 Build of semantic descriptions                         & 4 \\
 Build of the 3D visualization                          & 6 \\
 Architecture design                                    & 2 \\ 
 Implementation of customer services                    & 10\\ 
 Deployment in production                               & 2 \\
 \hline
\end{tabular}
\caption{Duration of stages for the proposed scenario}
\label{table:schedule}
\end{table}

\subsection{Build of semantic descriptions}
Given that the main products marketed by Urola Solutions are extruders, an ontology should be found that correctly describes this type of industrial machines. For this, it is necessary to define competency questions that describe this type of machines and  represent the questions that customers can ask about them when using the new services.
Below, some of the competency questions defined and categorized according to the service to which they correspond are presented.

\begin{enumerate}
    \item Catalogue
    \begin{enumerate}
        \item How many models of extruders does the company offer?
        \item What kind of product do these extruders make?
        \item What are the characteristics of the products manufactured by the extruders?
        \item Is there a 3D model of the extruder that can be visualized?
        \item What is the production (batch size) of a specific extruder model?
        \item[] ...
    \end{enumerate}
    \item Searching module
    \begin{enumerate}
        \item What is the volume of the bottle that an extruder produces?
        \item What is the size of the bottle?
        \item How many bottles per hour does an extruder produce?
        \item What is the necessary space to house an extruder?
        \item[] ...
    \end{enumerate}
    \item Virtual technician
    \begin{enumerate}
        \item What are the possible solutions for a problem with the motor?
        \item Where is the screw located?
        \item Which supplier has a compatible replacement fan?
        \item The extruder has stopped suddenly, what are the steps to follow?
        \item[] ...
    \end{enumerate}
\end{enumerate}

After defining the competency questions, it proceeded to look for an ontology that would describe this type of industrial machines (extruders) in the repositories indicated in the methodology. However, the search in these repositories did not give a favorable result, so a search was made with the words ``Extruder ontology" using google scholar. This search returned a scientific paper \cite{ramirez2020} with the description of an ontology named ExtruOnt.

As mentioned in \cite{ramirez2020}, the ExtruOnt ontology\footnote{\url{http://bdi.si.ehu.es/bdi/ontologies/ExtruOnt}} is the one that describes extruders most thoroughly. The ExtruOnt ontology represents different aspects related to extrusion machines. It includes terms to describe 1) the \textit{main components} of an extruder (e.g. the drive system), 2) the \textit{spatial connections} between the extruder components (e.g. the filter is externally connected to the barrel), 3) the \textit{different features} of the components (e.g. the power consumption of the motor is 40.5 kWh), 4) the \textit{3D description} of the position of the components (e.g. the feed hopper is located at point \textit{q(0,0,-1)} in a 3D canvas), and, 5) the \textit{sensors} that need to be used to capture indicators about the performance of that extruder (e.g the temperature sensor that captures the melting temperature of the polymer). This ontology has been implemented using OWL 2\footnote{\url{https://www.w3.org/TR/owl2-overview/}} and the Prot\'eg\'e\footnote{\url{https://protege.stanford.edu/}} development environment. Moreover, if an adequate ontology for another manufacturing scenario is not found, the detailed semantic description of the extruder that is made in the ExtruOnt ontology can serve as a model when making semantic descriptions of other products. 

Next, the ExtruOnt ontology was evaluated against the competency questions. Based on the classes (concepts) and relationships described in the ontology, the competency questions were represented using the SPARQL language, dummy individuals were created and the questions were executed, verifying that the obtained result was equal to the expected result. Thus, it was possible to validate that the extruder models offered by Urola Solutions could be correctly described using the ExtruOnt ontology and the competency questions could be fully answered. As an example, the SPARQL query for the competency question 1(e) is presented as follows:

\begin{Verbatim}[fontsize=\small]
PREFIX : <http://bdi.si.ehu.es/bdi/ontologies/ExtruOnt/Extruder01#>
PREFIX rdf: <http://www.w3.org/1999/02/22-rdf-syntax-ns#>
PREFIX owl: <http://www.w3.org/2002/07/owl#>
PREFIX s4inma: <https://w3id.org/def/saref4inma#>
PREFIX om: <http://www.ontology-of-units-of-measure.org/resource/om-2/>
PREFIX dcterms: <http://purl.org/dc/terms/>
SELECT ?batch ?size ?phenomenon ?description ?value
        WHERE {
            :E01 a ?restriction.
            ?restriction a owl:Restriction;
                owl:onProperty ?property;
                owl:allValuesFrom ?allValues.
            ?property owl:inverseOf s4inma:needsEquipment.
            ?allValues owl:intersectionOf ?intersection.
            ?intersection rdf:first ?batch;
                rdf:rest*/rdf:first ?node.
            ?batch rdfs:subClassOf s4inma:ItemBatch.
            ?node owl:hasValue ?size.
            ?size om:hasPhenomenon ?phenomenon.
            ?phenomenon dcterms:description ?description;
                om:hasNumericalValue ?value.
        }
\end{Verbatim}

After selecting the ontology, it proceeded to choose the knowledge platform. Due to the fact that the project deadlines were quite tight, it was decided to support the selection activity in evaluations that have already been carried out. Therefore, three different RDF stores were selected taking into account the evaluation carried out in \cite{addlesee_2019} and \cite{addlesee_2019b}: Virtuoso, Stardog and RDFox. For the evaluation of these RDF stores, the SPARQL queries resulting from the competency questions in the previous stage were used. Although these three RDF stores provide similar response times for most of the queries, Stardog presents too high response times for specific types of queries and does not offer  a free version, reasons why it was discarded. On the other hand, RDFox is an in-memory RDF store, which presents a great disadvantage in terms of data persistence, that is, if there is a crash or a system restart, the modifications made would be lost. The use of this RDF store involves creating and maintaining a backup system, which must respond to any eventuality. This, added to the fact that neither it offers a free version, has made this option be ruled out. Finally, Virtuoso is a hybrid database engine that combines the functionality of different types of databases in a single system. Virtuoso has fairly low response times ($<$500ms). Moreover, it offers an open-source version and  a SPARQL\footnote{https://www.w3.org/TR/sparql11-overview/} endpoint for connection from external systems, therefore it was the RDF store selected. 

Finally, an instance of Virtuoso Open-Source edition version 7.2.5 was installed in a Google Cloud virtual machine, in which the ExtruOnt ontology was loaded and the necessary namespaces and prefixes were defined. In addition, the operation of the endpoint was checked and the firewall rules for accepting requests were added. Furthermore, the SPARQL queries were executed again to check the correct functionality and performance of the knowledge platform.

\subsection{Build of the 3D visualization} 
As a product manufacturer, Urola Solutions owns their CAD models. These models were created using a desktop CAD application, whose license is quite expensive and for which the period of free updates had already expired, and in addition it does not support an export method other than saving these models on the local disk. Within the objectives of this phase it was decided to look for an alternative CAD application that offered additional advantages with respect to the current CAD application as well as an export method according to the needs of the project.

Among the different CAD packages found for exporting CAD models, those based on cloud processing 
(Onshape, Autodesk Fusion 360, etc.) 
caught our attention for the reasons explained in section \ref{section:build3dvisualization}.
Among those options, it was decided to use Onshape since it has a free plan with which all the functionalities can be tested without incurring in any cost, it also allows collaboration between teams from mobile and desktop devices and has an API which can be used for exporting CAD models to the user application.

Using two CAD models of extruders provided by Urola solutions, it was verified that these models were fully customizable within the Onshape application and that the connection to the Onshape API for the export will work correctly. It should be noted that Onshape provides a client repository for its API\footnote{\url{https://github.com/onshape-public}} in different programming languages and runtime environments, one of them for NodeJS, which was the one used considering that the software engineers involved in the project had previous experience developing on it.

\begin{figure}[t] 
\centering
\includegraphics[width=0.7\columnwidth]{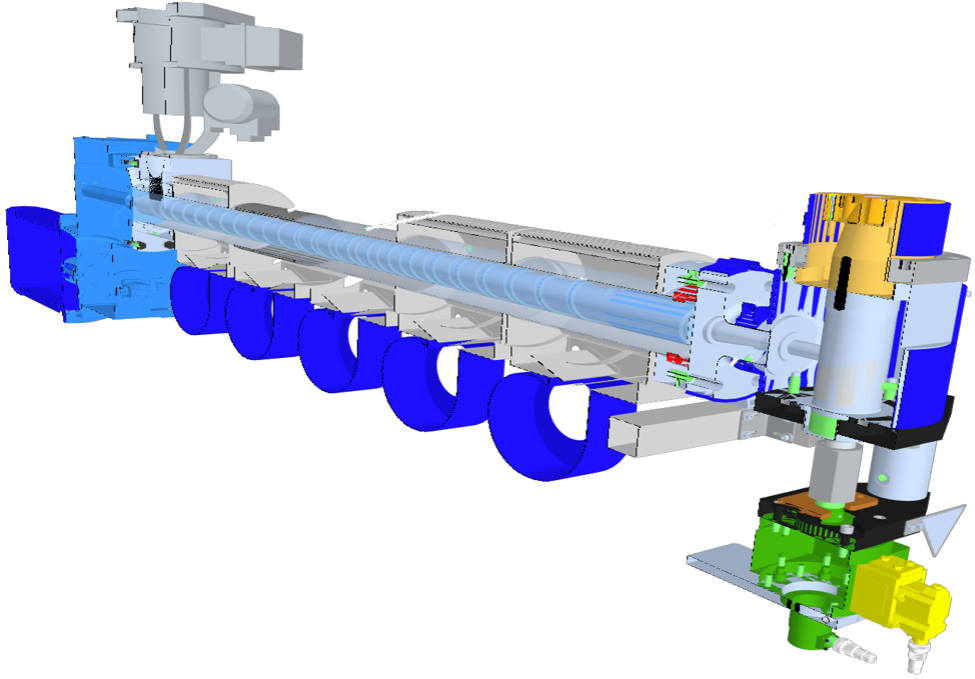}
  \caption{\label{fig:3DRender}
     3D extruder visualization on the rendering interface.}
      \end{figure}

The fact that the new services would be available to the client through web browsers constituted a compelling reason for the selection of the 3D platform, for this reason it was necessary to look for frameworks or libraries that would support the rendering of CAD models through this medium. There are different open-source frameworks based on WebGL (an OpenGL based javascript library) for 3D rendering in web browsers. Among these, XML3D, X3DOM and Three.js stand out \cite{EVANS201443}. The later one was selected because it has a simple learning curve, extensive documentation, lots of tutorials, and a powerful community. Using the selected 3D rendering framework, an interface was designed and built for the visualization of the extruder CAD models. The interface allows to visualize and interact with the CAD models of the extruders using actions controlled with the mouse and keyboard, this includes moving, rotating, zooming in, zooming out, selecting, making cuts to the model, etc. Figure \ref{fig:3DRender} shows an example of the visualization of an extruder with a vertical cut in the rendering interface.

The last activity was to perform communication tests between the Onshape API and a test application that incorporated the rendering interface in order to check the compatibility, synchronization and visualization of CAD models exported from Onshape. Those tests gave satisfactory results.

\subsection{Architecture design}
Following the methodology activities for this stage, the hardware resources available to configure the architecture were identified. Urola Solutions has outsourced its infrastructure, so any inclusion of resources to this infrastructure would generate an impact on the budgeted expense. 
For this reason, the company has a development environment on Google Cloud for the proof of concept of its new projects, which includes more than 20 free products (with a monthly usage limit) such as: Firebase, Firestore, Compute Engine, Cloud Storage, App Engine, among others. Therefore, it was decided to deploy the architecture on this development environment.

From the different types of architecture considered, the 3-Tier architecture type was the one that best suited the needs of the project, since in the future a native client for mobile devices could be developed, so it was necessary to separate the presentation tier from the application tier. 



For the presentation tier of the selected architecture it was decided to use React \footnote{\url{https://reactjs.org/}}, a javascript library created by Facebook for the construction of user interfaces, with which the software engineers of the project had experience enough which would shorten programming times. In addition, FireBase \footnote{\url{https://firebase.google.com/}} was used as hosting service since it has tools that facilitate administration, user management and online testing. FireBase also provides a free version (Spark plan) with more than enough resources for the development of the application, such as 1 GB of database storage in Cloud Firestore, 10 GB of hosting storage, custom domains and SSL security.

For the application tier, a REST API in charge of the functional business logic and the communication between the presentation tier and the data tier was developed using NodeJS \footnote{\url{https://nodejs.org/}}, an execution environment that allows to use javascript code in server; and ExpressJS \footnote{\url{https://expressjs.com/}}, a web application framework for NodeJS designed for the creation of APIs among others.

\begin{figure} 
\centering
\includegraphics[width=\columnwidth]{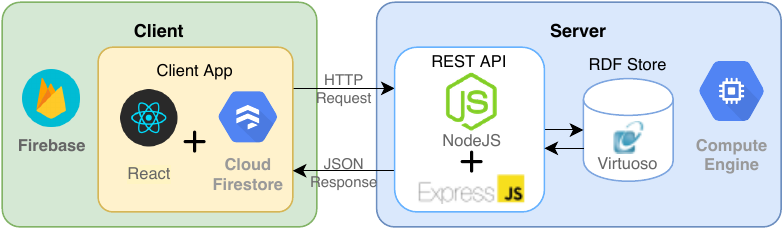}
  \caption{\label{fig:architecture}
     Architecture.}
      \end{figure}

For the data tier, a virtual machine instance was created in the Google Cloud Compute Engine, specifically of type n1-standard-1 (1 virtual CPU, 3.75 GB of memory) with a hard disk of 30 GB, in which the knowledge platform was deployed. It was also decided to use the same virtual machine instance to host the API (application tier), in order to emulate the production deployment of the project since they would share the same server.
Figure \ref{fig:architecture} shows the main blocks of the designed architecture. The architecture was tested configuring one user in the Cloud Firestore database and making calls to the API, using a demo React app, to query the available extruders in the knowledge platform. Security configurations were applied to the virtual machine instance (application and data tiers) in order to only accept incoming requests from the React App deployed in Firebase (presentation tier).

\subsection{Implementation of customer services}
Three services that guarantee an improvement in the customer life cycle and that integrate the technologies selected in the previous stages have been implemented. These are a catalogue, a searching module and a virtual technician. The first two services are related to the Discover \& Shop  phase and the last one is related to the Use \& Service phase of the customer life cycle.

Before proceeding with the design of the services, an analysis of similar services was made as stated in the methodology. The development of this activity gave us important guidelines on how the new services should be approached taking into account the characteristics, advantages and shortcomings of the service types analyzed, which are presented next. 

Regarding the catalogues, on the one hand, different types of catalogues are used on e-commerce platforms (Business-to-Costumer (B2C) model). They are developed to  improve the end customer shopping experience by offering intuitive visualizations  and a recommendation system based on previous search and purchase preferences.
On the other hand, in the Industry 4.0 environment, most of the companies  provide their customers with a generic physical or digital brochure with a lack of depth description of the goods or services offered.
Taking into account that many times their customers are other businesses, i.e. Business-to-Business (B2B) model, those brochures are insufficient because they do not incorporate many technical details and thus, the purchasing processes are carried out in extensive meetings between the purchasing and sales departments of each company, taking more time than actually necessary.
A great advance in this aspect is introduced with the use of an online 3D rendering technology in the developed catalogue. Thus, a customer can see first-hand what the desired product looks like and moreover, the visualization is enriched with relevant information in such a way that the user experience is greatly improved. 

Regarding the search modules, most of them are inserted into the catalogues and its operation is based on keywords, which can be a great challenge for novice users who do not know the correct terminology to perform a search (i.e. exact name, model, serial number, functionality, etc). To avoid this limitation, the search module developed is initially based on a battery of simple questions that guide the customer in the selection process.

\begin{figure}[t] 
\centering
\includegraphics[width=\columnwidth]{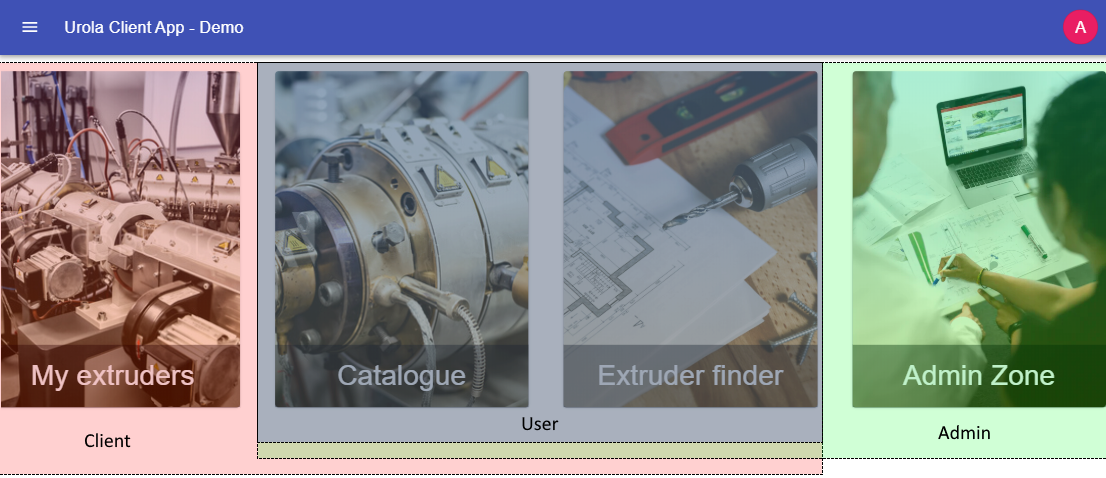}
  \caption{\label{fig:HomeScreen}
     Home screen options in the user application.}
      \end{figure}

Regarding post-purchase services, the most common way of providing a post-purchase service that includes repair, maintenance and support is through the customer service lines. However, this solution presents different problems such as fixed service hours, congestion on the lines, relatively long waiting times and the high possibility of not finding an immediate solution. Those problems can generate high customer discomfort creating an effect contrary to the desired loyalty. The effect is much more damaging in the manufacturing industry, where enormous economic loss can be generated by not having the requested spare part or not having carried out preventive and corrective maintenance on time. In order to avoid those problems the developed service (virtual technician) provides a troubleshooting module available 24 hours a day with the possibility of requesting spare parts directly from the main supplier or from other suppliers. 

With the information resulting from the analysis of similar services, we proceeded to the design and code writing of the new services.
However, before building those services, the REST API and the user application where the services will be hosted must be developed.
 Regarding the user application, it must contain an administration module (Admin Zone in figure \ref{fig:HomeScreen}) in which the information that will be available for the services will be managed. Figure \ref{fig:HomeScreen} shows the home screen of the user application with the administration module and the available services, as well as the roles that can access them. 
	  
\begin{figure}[t] 
\centering
\includegraphics[width=\columnwidth]{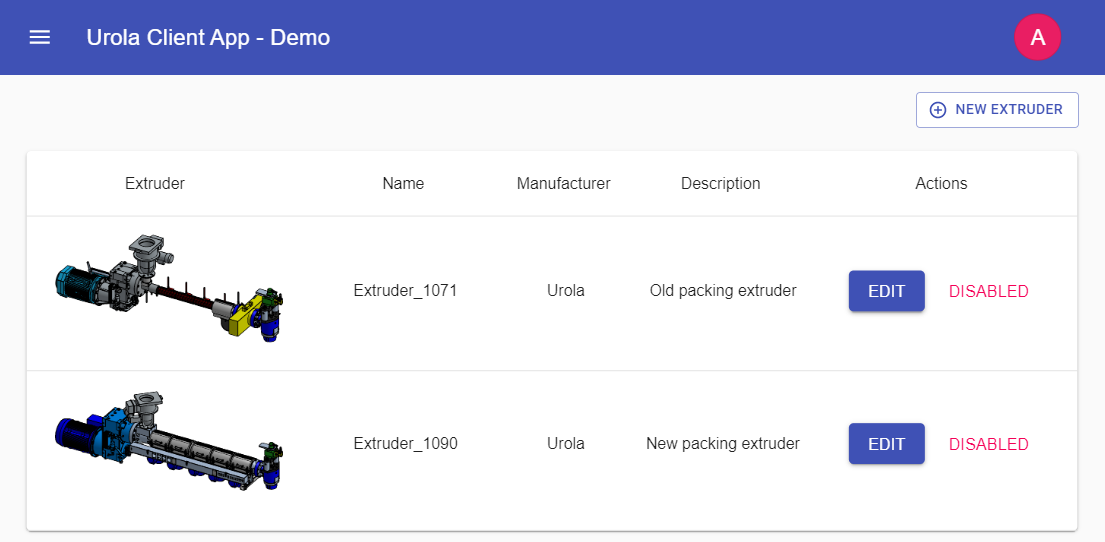}
  \caption{\label{fig:ExtruderManager}
     Main screen of administration module.}
      \end{figure}

In the administration module, the necessary annotations are generated, using the descriptions of the ExtruOnt ontology, to store the information of the extruder models in the knowledge platform. This module is accessible only to users with an administrator role. The first screen of this module (Figure \ref{fig:ExtruderManager}) presents a list of the extruder models already loaded, with the option to edit or disable their display, and a button to access a sub-module where a new extruder model can be created. This sub-module presents a dynamic form where the main characteristics of an extruder model (name, manufacturer, description, etc.) can be specified, in addition to the list of its components and their characteristics. It should be noted that the necessary information to create the form was completely extracted from the descriptions present in the ontology, with which the generation of a database schema from scratch was not necessary, being this one of the advantages of ontologies. 

\begin{figure}[t] 
\centering
\includegraphics[width=\columnwidth]{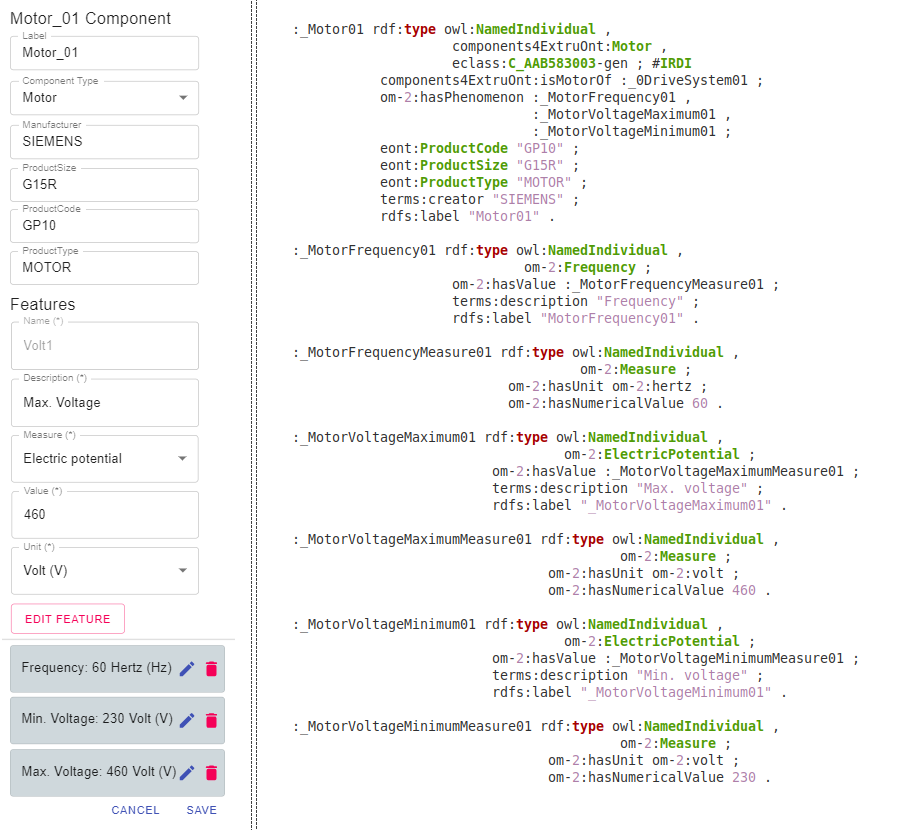}
  \caption{\label{fig:FeaturesRDF}
     Annotations generated when describing motor features.}
      \end{figure}

The left side of Figure \ref{fig:FeaturesRDF} presents an example of the form for describing a component of an extruder, more precisely its motor. The field \textit{Component Type} links this instance of motor to the class \texttt{Motor} in ExtruOnt. Thanks to the information available in ExtruOnt about \texttt{Motor}s, the \textit{Features} section of the form is dynamically personalized to allow only for measure types that can be applied to motors (e.g. Electric potential, Frequency). In the same way, once a measure type is selected (e.g. Electric potential) the field \textit{Unit} is filled with appropriate units for that type of measure with regard to the information in ExtruOnt. The user can add as many as features as they want about the selected component (e.g. a frequency of 60Hz, a minimum voltage of 230V, a maximum voltage of 460V), and then this information is internally transformed into a set of RDF triples annotated with the classes and properties of ExtruOnt, as can be seen in the right side of Figure \ref{fig:FeaturesRDF}. Moreover, apart from the triples generated from the information available in the form, new triples that link the instance with additional relevant information inherited from the type of the component are generated. One example of additional information is the International Registration Data Identifier (IRDI) code of the type of component, whose semantic representation is available through the eClassOWL ontology\footnote{http://www.heppnetz.de/projects/eclassowl/}. This ontology models eCl@ss\footnote{https://www.eclass.eu}, a classification standard for products, and is used to facilitate interoperability.

\begin{figure}[] 
\centering
\includegraphics[width=0.9\columnwidth]{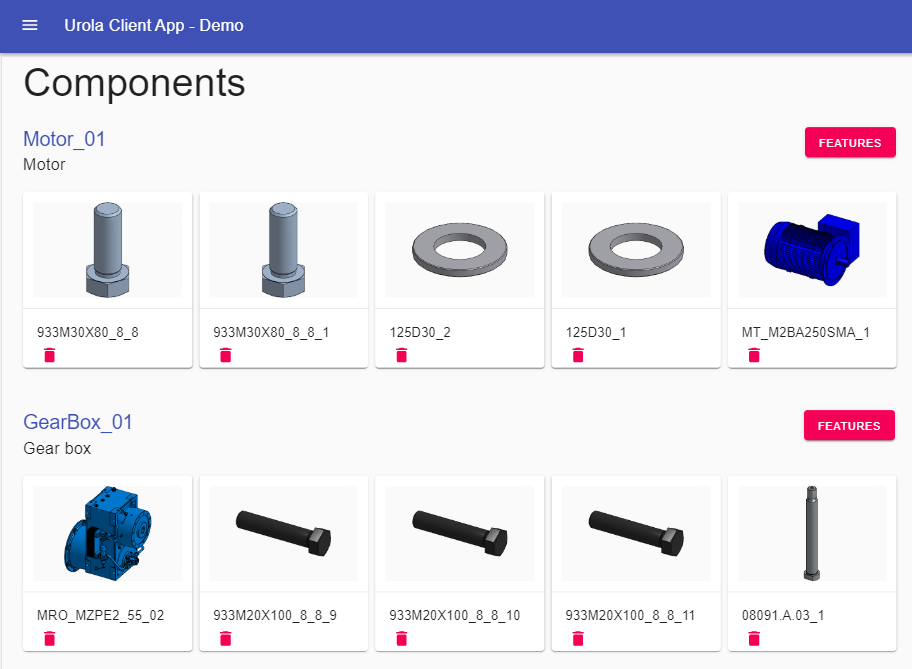}
  \caption{\label{fig:ClassificationSystem}
     Import system.}
      \end{figure}

\begin{figure}[] 
\centering
\includegraphics[width=0.9\columnwidth]{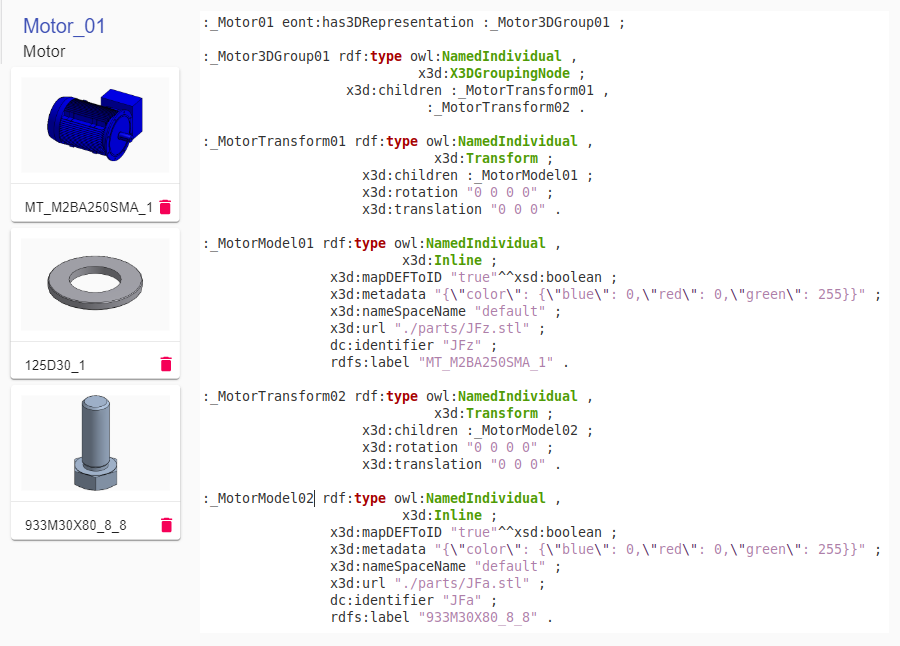}
  \caption{\label{fig:3DAnnotations}
     Annotations generated when describing motor 3D models.}
      \end{figure}

Additionally, this sub-module also includes a section where CAD models can be imported from the Onshape API and whose information is annotated in the knowledge platform using the descriptions of the 3D module from ExtruOnt. 
Figures \ref{fig:ClassificationSystem} and \ref{fig:3DAnnotations} show an example of the implemented import system and the generated annotations respectively.
In the following the functionality of each of the services is described.

\subsubsection{ Catalogue}

\begin{figure}[] 
\centering
\includegraphics[width=0.9\columnwidth]{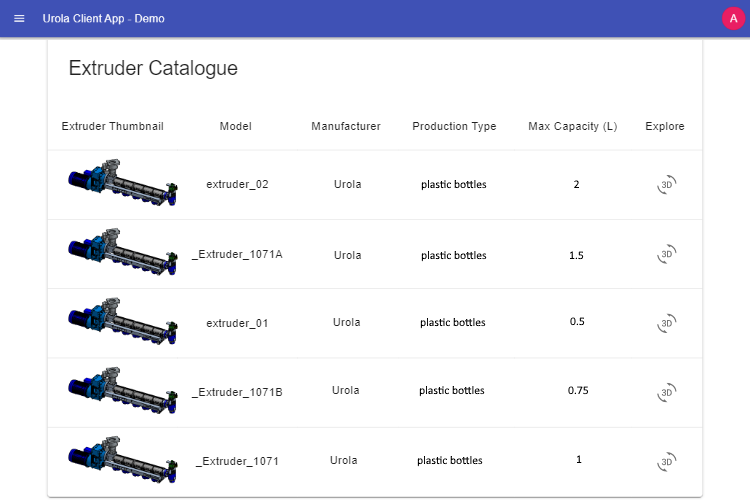}
  \caption{\label{fig:CatalogueMainScreen}
     Main screen of the Catalogue.}
      \end{figure}
      
\begin{algorithm}
    \footnotesize
    \caption{Get all extruders with their components}\label{alg:catalogue}
    \hspace*{\algorithmicindent} \textbf{Input: } $\emptyset$ \\
    \hspace*{\algorithmicindent} \textbf{Output:} \( \mathcal{R} \): Object list with extruders, components, properties and 3D models.\\
    \hspace*{\algorithmicindent} \textbf{Variables: } \( \mathcal{E} \): Object list containing extruders with their main properties.\\
    \hspace*{2.2cm} \( \mathcal{P} \): Object list containing extruder components.\\
    \hspace*{2.2cm} \( \mathcal{C} \):  Object list containing extruder components with their properties and 3D models.
    \begin{algorithmic}[1]
    \Function{GetAllExtruders}{}
    \State $\mathcal{R} \gets \{\}$
    \State $\mathcal{E}  \gets \textit{getDataS(``allExtrudersList") }$\Comment{Run SPARQL query named ``allExtruderList"}
    \ForEach {$e \in \mathcal{E} $}
    \If{$e_{visible} = $ true } \Comment{Only visible extruders}
    \State $e_{parts} \gets partsById(e_{id})$ \Comment{Inject the extruder components}
    \State append $\langle e \rangle$ to $\mathcal{R}$ \Comment{Append extruder to list}
    \EndIf
    \EndFor
    \State \Return $\mathcal{R}$
    \EndFunction
    \Function{partsById}{$e_id$}
    \State $\mathcal{C} \gets \{\}$
    \State $\mathcal{P}  \gets \textit{getDataS(``partsByExtruderId") }$
    \ForEach {$p \in \mathcal{P} $}
    \State $p_{properties} \gets propertiesById(p_{id})$ \Comment{Inject the component properties}
    \State $p_{model} \gets modelsById(p_{id})$ \Comment{Inject the component 3D models}
    \State append $\langle p \rangle$ to $\mathcal{C}$ \Comment{Append component to list}
    \EndFor
    \State \Return $\mathcal{C}$
    \EndFunction
    \end{algorithmic}
\end{algorithm}

\begin{figure}[] 
\centering
\includegraphics[width=0.8\columnwidth]{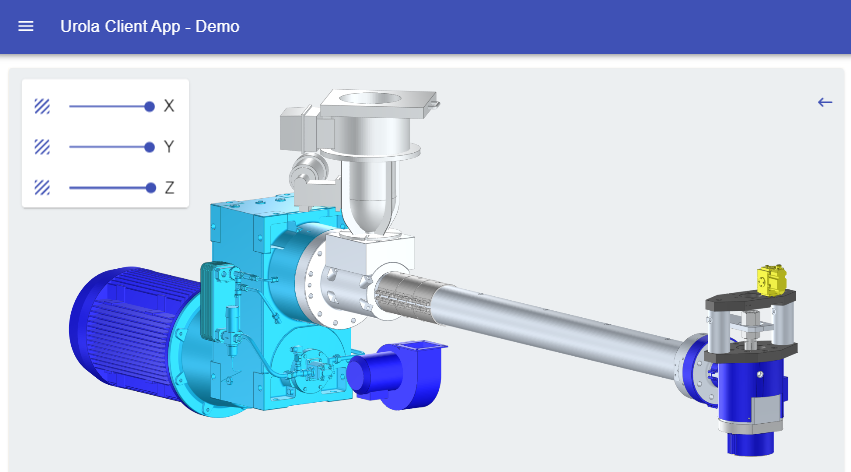}
  \caption{\label{fig:3DCatalogue}
     3D extruder rendered in a canvas.}
      \end{figure}

The developed catalogue 
(Figure \ref{fig:CatalogueMainScreen}) is made up of a list of the different extruder models available together with the information corresponding to each model, manufacturer and production, as well as a navigation button which leads to 3D visualization. 
Algorithm \ref{alg:catalogue} describes the process to obtain all extruders with their components, properties and 3D models inside the catalogue load workflow.
This information is obtained from the annotated data in the knowledge platform managed by the administration module and rendered in a canvas using the Three.js framework (Figure \ref{fig:3DCatalogue}). The interaction with visualization and navigation is carried out using the mouse to control the events of the scene, in this way it is possible to zoom in, zoom out and rotate the 3D model. It is also possible to make cuts to the model in the three dimensional axes using the sliders in the upper left corner of the screen, allowing to view and select components that were hidden. The selection of the components of the extruder displays relevant information for the user related to that component such as type, model and brand. There is also a button to request more information, with which a contact form is displayed. Its information contains, in addition to the data filled in by the potential client, the information of the extruder model displayed, helping the sales representative to better guide the first contact with the user in order to make them a future customer. 

\subsubsection{Searching module}
\begin{figure} 
\centering
\includegraphics[width=\columnwidth]{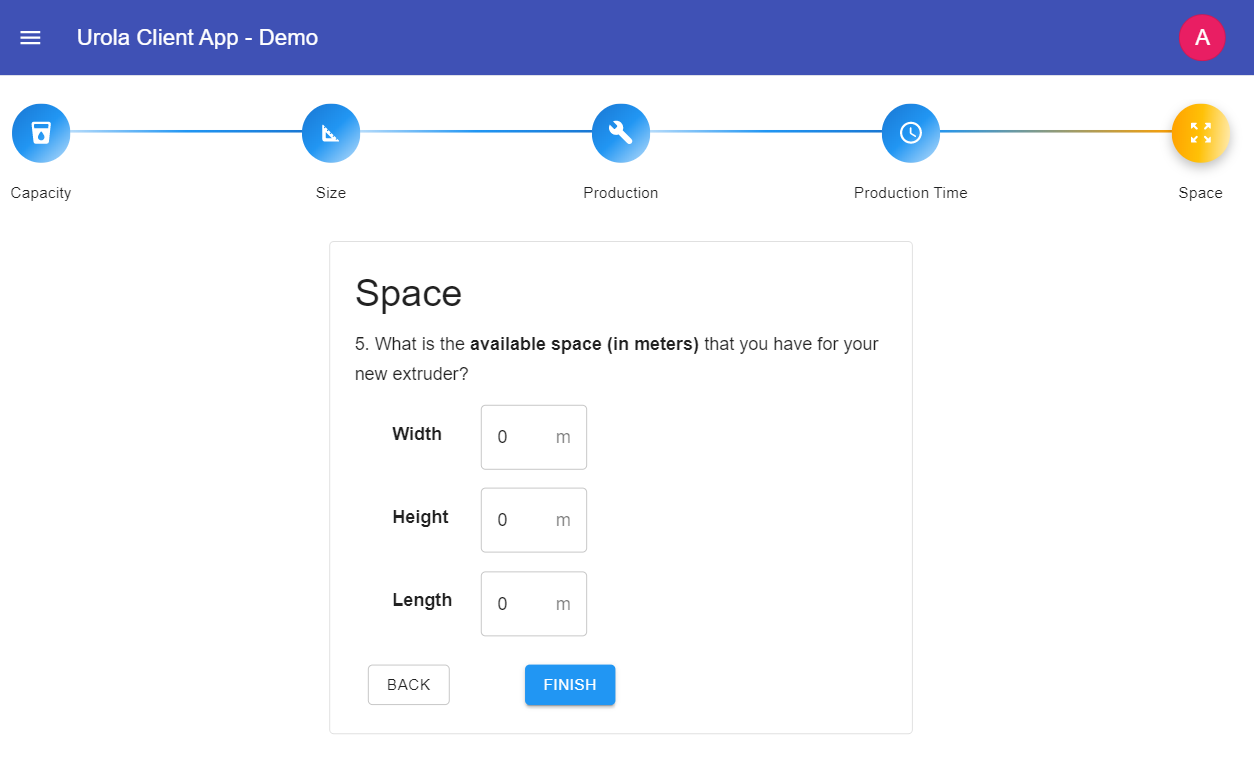}
  \caption{\label{fig:Questions}
     Example of a question in the searching module.}
      \end{figure}

The searching module presents two types of interfaces: simple and advanced.
For the simple type, five general questions are presented, focused on the production and dimension of the extruders, which help customers to find the ideal product according to their requirements. Based on the answers to the questions, a SPARQL query will be formulated, which will be executed against the annotated information about the different extruder models. These questions are:

\begin{itemize}
    \item What is the volume of the bottle that you want to produce?
    \item What is the size of the bottle?
    \item How many bottles would you like to produce?
    \item How many hours does your company work per day?
    \item What is the available space (in meters) that you have for your new extruder? (see Figure \ref{fig:Questions})
\end{itemize}

\begin{figure} 
\centering
\includegraphics[width=0.93\columnwidth]{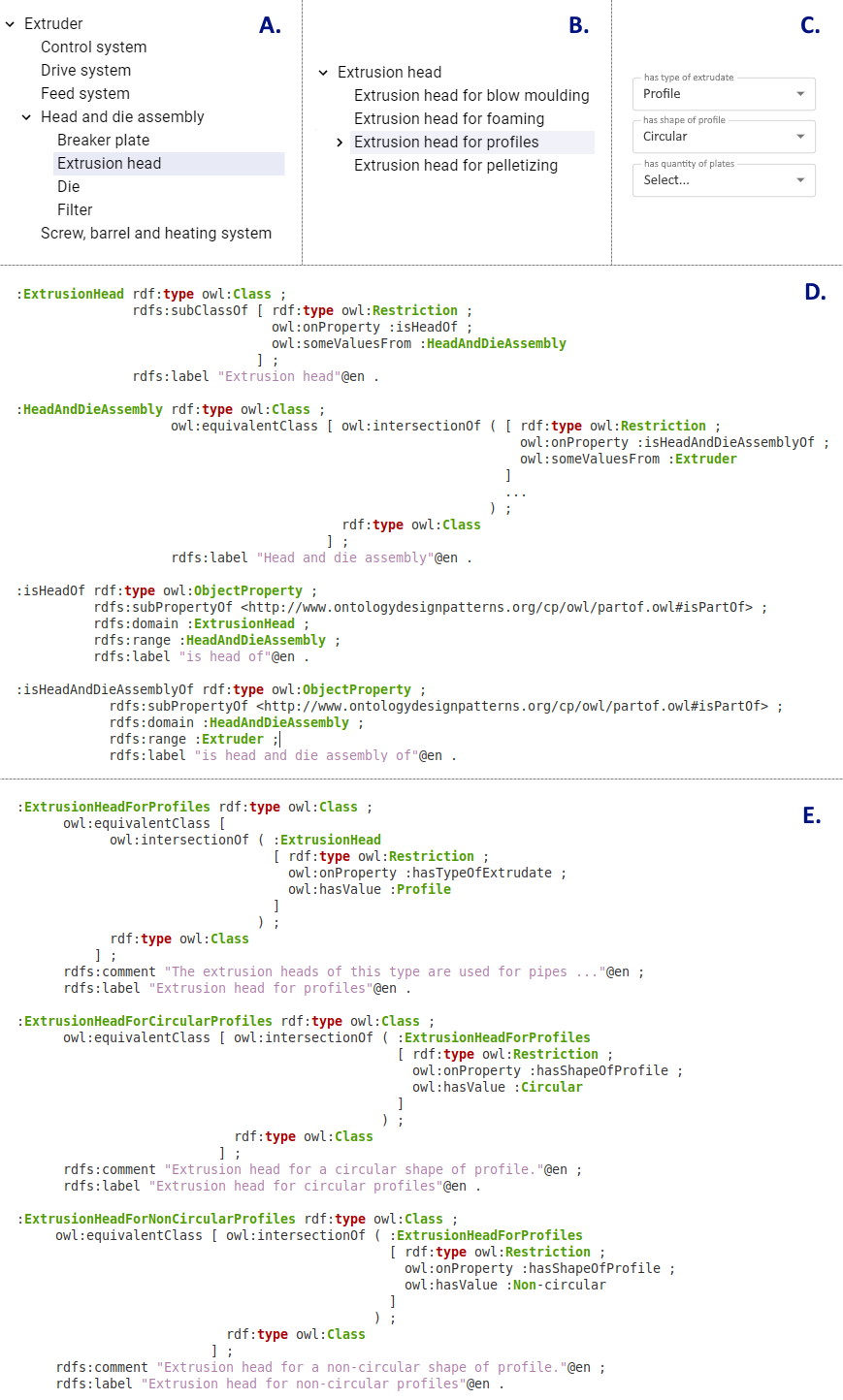}
  \caption{\label{fig:SearchingModule}
     Example of annotations used to construct the advanced interface of the searching module.}
      \end{figure}

Algorithm \ref{alg:searchingmodule} describes the procedure carried out when running a basic search.

\begin{algorithm}
    \footnotesize
    \caption{Get all extruders with basic search parameters}\label{alg:searchingmodule}
    \hspace*{\algorithmicindent} \textbf{Input: } \( \mathcal{P}_{vol} \): Volume of the bottle.\\
    \hspace*{1.65cm} \( \mathcal{P}_{wh} \): Object with width and height of the bottle.\\
    \hspace*{1.65cm} \( \mathcal{P}_{prod} \): Quantity of bottles to produce in a day.\\
    \hspace*{1.65cm} \( \mathcal{P}_{hpd} \):  Working hours per day.\\
    \hspace*{1.65cm} \( \mathcal{P}_{esize} \):  Object with width, height and length of the extruder.\\
    \hspace*{\algorithmicindent} \textbf{Output:} \( \mathcal{R} \): Object list with extruders, components, properties and 3D models.\\
    \hspace*{\algorithmicindent} \textbf{Variables: } \( \mathcal{E} \): Object list containing extruders with their main properties.\\
    \hspace*{2.2cm} \( \mathcal{F} \): String with filter options.
    \begin{algorithmic}[1]
    \Function{GetAllExtrudersByParams}{$\mathcal{P}_{vol},\mathcal{P}_{wh},\mathcal{P}_{prod},\mathcal{P}_{hpd},\mathcal{P}_{esize}$}
    \State $\mathcal{R} \gets \{\}$
    \State $\mathcal{Q}_{sparql} \gets \textit{getQuery(``BasicSearchQuery") }$ \Comment{Return query template for basic search}
    \State $\mathcal{F} \gets \textit{ValidateFilters(} \mathcal{P}_{vol},\mathcal{P}_{wh},\mathcal{P}_{prod},\mathcal{P}_{hpd},\mathcal{P}_{esize} \textit{)}$ \Comment{Return filter options}
    \State append $\langle \mathcal{F} \rangle$ to  $\mathcal{Q}_{sparql}$ \Comment{Append filter options to query}
    \State $\mathcal{E}  \gets \textit{getDataQ(}\mathcal{Q}_{sparql}\textit{)}$\Comment{Run SPARQL query with filter options}
    \ForEach {$e \in \mathcal{E} $}
    \If{$e_{visible} = $ true } \Comment{Only visible extruders}
    \State $e_{parts} \gets partsById(e_{id})$ \Comment{Inject the extruder components}
    \State append $\langle e \rangle$ to $\mathcal{R}$ \Comment{Append extruder to list}
    \EndIf
    \EndFor
    \State \Return $\mathcal{R}$
    \EndFunction
    \end{algorithmic}
\end{algorithm}

The advanced interface complements the previous one, allowing searches based on the components of the extruders, and taking advantage of the inheritance characteristics of the Web Ontology Language (OWL) with which the ExtruOnt ontology was built. More precisely, it allows to indicate the specific class to which a component must belong. This advanced search engine is activated in the option \textit{Add advanced condition} that is displayed after answering the questions of the simple search engine and consists of two sections. First, it shows the component tree of an extruder (Figure \ref{fig:SearchingModule}A), which is generated from the parthood relationships present in the ontology (Figure \ref{fig:SearchingModule}D). Selecting a component will display the specializations of that component (subclasses), allowing further refinement of the search (Figure \ref{fig:SearchingModule}B). 
Moreover, properties related to the selected specialization will be shown (Figure \ref{fig:SearchingModule}C). Each of the properties will serve for one of the following two purposes: information or refinement. On the one hand, properties for information purposes are those that according to ExtruOnt are either associated to the selected specialization or inherited from its superclasses. These properties will have a fixed value for that component. In the example shown in Figure \ref{fig:SearchingModule}, due to the description of the class \texttt{ExtrusionHeadForProfiles} in ExtruOnt (Figure \ref{fig:SearchingModule}E), the property \textit{has type of extrudate} has been assigned the fixed value \textit{Profile}. On the other hand, as their name indicate, properties for refinement will help refine the search by allowing the user to provide values for them. More precisely, they will be those properties in ExtruOnt appearing in the restrictions of the subclasses of the selected specialization. In the example, a restriction exists in classes \texttt{ExtrusionHeadForCircularProfiles} and \texttt{ExtrusionHeadForNonCircularProfiles} which states values of \texttt{Circular} and \texttt{Non-circular} respectively for property \texttt{hasShapeOfProfile}. Thus, those two values will appear for refinement in the property \textit{has shape of profile} of the form. Although the related annotations are not shown for space matters, the same applies for the property \textit{has quantity of plates}. 

Once the search has been carried out, simple or advanced, a list of those extruders that meet the indicated conditions are displayed using the same format and functionality of the catalogue. It is worth mentioning that when the form to request more information is used on an extruder obtained through the search module instead of the catalogue, the search parameters are included in the information sent to the sales representative.

\subsubsection{Virtual technician}

\begin{figure}[t] 
\centering
\includegraphics[width=\columnwidth]{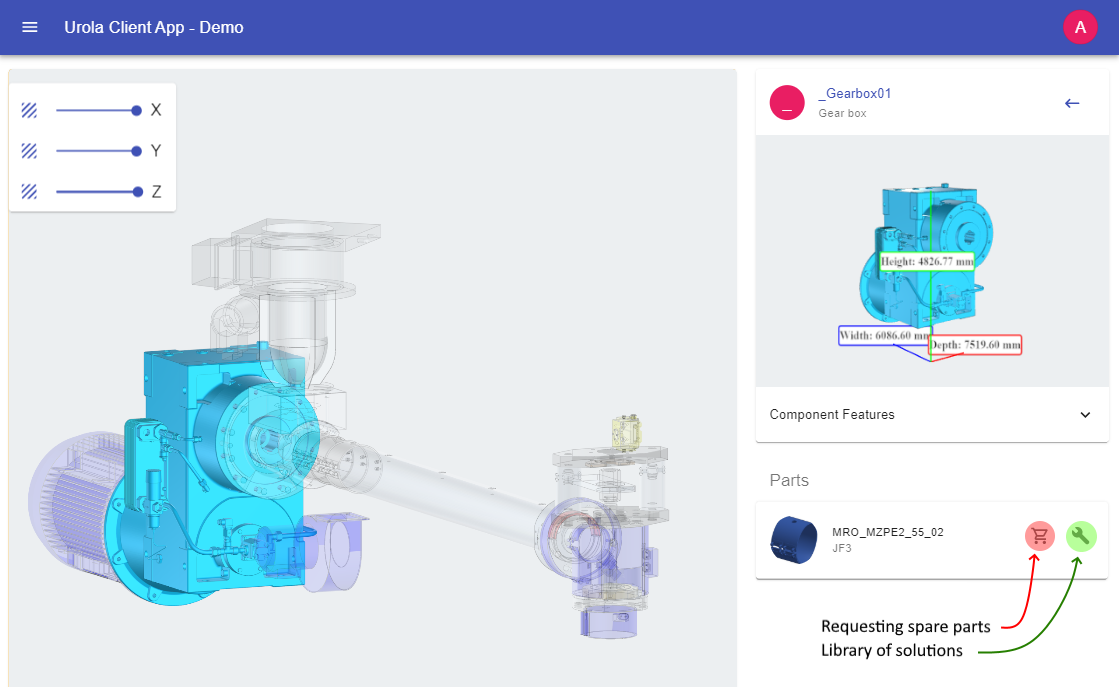}
  \caption{\label{fig:MyExtruders}
     Virtual technician: Customer options after selecting an extruder component.}
      \end{figure}

Using this service, customers can see the list of their extruders bought or rented (B2C or B2B model, respectively) with the same format used for the catalogue and with the same interaction capacity in the 3D visualization (Figure \ref{fig:MyExtruders}). Technically it is supported by 
two modules: 1) A library of solutions for the most common problems generated by the manipulation of extruders created by Urola Solutions from the experience gained over the years. This library is annotated in the knowledge platform and associated with the extruder components through the custom property ``related to" whose domain and range are a problem and a component, respectively. The system loads the library by filtering through the extruder component that is selected in the 3D visualization, avoiding showing irrelevant information and guiding the user step by step in solving the problem. If there is no solution, a support ticket is opened with which the technician will receive the history of previous actions, so that the client does not have to explain the problem from scratch. 2) A module for requesting spare parts in which individual parts can be requested from the main supplier. In case of not having a stock of parts, other suppliers where stock is available are suggested. The query of the available stock and other suppliers is done through the already-existing internal company service using its own codes for the components. However, when needed, IRDI codes in the annotations of the components could be used to broaden the search to other components classified under the same IRDI code. 
Algorithm \ref{alg:virtualtechnician} describes the procedure carried out when a spare part is requested.
 
\begin{algorithm}
    \footnotesize
    \caption{Requesting spare part to available provider}\label{alg:virtualtechnician}
    \hspace*{\algorithmicindent} \textbf{Input: } \( \mathcal{C}_{id} \): Component identifier.\\
    \hspace*{1.65cm} \( \mathcal{P}_{id} \): Provider identifier.\\
    \hspace*{\algorithmicindent} \textbf{Output:} \( \mathcal{R} \): Response object with provider list or order status.\\
    \hspace*{\algorithmicindent} \textbf{Variables: } \( \mathcal{A} \): Part availability.\\
    \hspace*{2.2cm} \( \mathcal{L} \): List of alternative providers.\\
    \hspace*{2.2cm} \( \mathcal{O} \): Order status.
    \begin{algorithmic}[1]
    \Function{RequestSparePart}{$\mathcal{C}_{id},\mathcal{P}_{id}$}
    \State $\mathcal{R} \gets \{\}$
    \State $\mathcal{A} \gets partAvailableInWarehouse(\mathcal{C}_{id})$ \Comment{Always check availability in warehouse first}
    \If{$\mathcal{A} = true$ } 
        \State $\mathcal{O} \gets requestPartToWarehouse(\mathcal{C}_{id})$
        \State $\mathcal{R} \gets \langle ``Order",\mathcal{O} \rangle$ \Comment{Return order status}
    \ElsIf{$\mathcal{A} = false \textbf{ and } \mathcal{P}_{id} = \emptyset$ } 
        \State $\mathcal{L} \gets GetProvidersByPartId(\mathcal{C}_{id})$ \Comment{Get list of alternative providers or NULL}
        \State $\mathcal{R} \gets \langle ``Providers",\mathcal{L} \rangle$ \Comment{Return available providers}
    \Else
        \State $\mathcal{L} \gets GetProvidersByPartId(\mathcal{C}_{id})$
        \If{$\mathcal{P}_{id} \textbf{ included in } \mathcal{L}$}
            \State $\mathcal{O} \gets requestPartToProvider(\mathcal{C}_{id},\mathcal{P}_{id})$
            \State $\mathcal{R} \gets \langle ``Order",\mathcal{O} \rangle$
        \Else
            \State $\mathcal{R} \gets \langle ``Providers",\mathcal{L} \rangle$
        \EndIf
    \EndIf
    \State \Return $\mathcal{R}$
    \EndFunction
    \end{algorithmic}
\end{algorithm}

\vspace{3mm} 

At the time of writing this paper, the activities of usability analysis and testing of functionality, integration, interoperability and user acceptance are being carried out by Urola Solutions. However, the feedback of the 
customers that are trying the system is encouraging. They really value the possibilities it offers in relation to product exploration,  training of employees and recommendations regarding spare parts. Moreover, the results regarding the evaluation of the system response times carried out in each of the  developed services in those real manufacturing scenarios, varying the information load, can be seen in Table \ref{table:evaluation} (it shows the average times obtained for the most relevant tests). As  can be observed, most actions are below the recommended average time of 4.7 seconds for loading web pages\footnote{\url{https://www.machmetrics.com/speed-blog/average-page-load-times-for-2020}} even using the limited deployment environment that was available.\\

\begin{table}[h!]
\centering
\begin{tabular}{l c} 
 \hline
 Action & Avg. response time (s) \\ 
 \hline\hline
 Catalogue loading                                  & 0.938 \\ 
 Extruder insertion                                & 2.457 \\
 Loading the 3D rendering of an extruder                                 & 9.631 \\
 3D models data import from OnShape                                  & 2.736 \\
 Loading the library of solutions                                  & 1.428\\
 \hline
\end{tabular}
\caption{Average response times for system evaluation (in seconds)}
\label{table:evaluation}
\end{table}

\subsection{Deployment in production and maintenance.}
This stage and its activities will be addressed once the last activities of the previous stage have been completed.

\section{Conclusions and future work}


\label{conclussion}

This paper presents a novel approach focused on the customer life cycle to facilitate the implementation of Industry 4.0 in those Small and Medium Enterprises belonging to traditional manufacturing sectors, whose limited resources and the high degree of complexity that this entails, make them desist from starting this process.

The main contribution of this paper is a methodology that describes a series of well defined stages and activities, easy to understand and execute, with which this transition can be carried out using minimal economic resources and taking advantage of new technologies that were not previously easily accessible. This methodology is mainly based on the use of semantic technologies and 3D visualization, which have been extensively explored individually, but to the best of our knowledge, they have
not been used together.
On the one hand, semantic technologies, in this case ontologies, provide a high degree of flexibility for the description of knowledge, in addition to allowing inference and reasoning capabilities that are difficult to achieve by traditional databases. On the other hand, 3D rendering technologies offer an enhanced visual representation that includes better graphics and navigation controls, allowing the user an interactive and improved experience. All these benefits focused on a particular type of user, the customer, throughout their life cycle will improve the relationship between the customer and the enterprise, achieving a high degree of loyalty.

The second contribution is a system, developed as a proof of concept in a real manufacturing enterprise, in which the introduced methodology has been followed step by step and the previously described technologies have been used to generate a series of services that positively affect the relationship with the customer in two of the three phases of the customer life cycle. This system has an underlying ontology that allows to describe extrusion machines in a reliable and flexible way, making the proposal easily extensible to any company that works with this type of machine. Furthermore, the ontology can be modified to describe any other type of product (e.g. toys, furniture) and the system can be adapted to be used by other manufacturing enterprises. The development of this system, currently in the last stage of testing before being put into production, serves as an example of the effectiveness of the proposed methodology.

Future work contemplates the creation of a native application for mobile devices, improving the scope of the system. 
It is also important to explore the improvements that can be obtained in the second phase of the customer life cycle, i.e. Buy \& Install, eventually expanding the number of services offered with services related to channel partners (e.g. dealers or distributors), such as visualizing and analyzing real-time data from these partners for optimizing delivery, predicting issues and making better operational decisions.

\section*{Acknowledgements}
The authors wish to thank Urola Solutions for allowing us to carry out the proof of concept, for their help with information about the customer service process and for providing real data.

\bibliography{mybibfile}

\end{document}